\documentclass[12pt]{article}
\usepackage{epsfig}
\usepackage{amssymb,amsfonts}
\newcommand{\beq}{\begin{equation}}
\newcommand{\eeq}{\end{equation}}
\newcommand{\bea}{\begin{eqnarray}}
\newcommand{\eea}{\end{eqnarray}}

\begin{document}
\title{\bf Stability of strange quark matter:\\ MIT bag versus Color Dielectric Model}
\author{
W.M. Alberico$^{\mathrm{a}}$,
A. Drago$^{\mathrm{b}}$,
C. Ratti$^{\mathrm{a}}$
\vspace*{0.3cm}\\
\begin{tabular}{c}
{\it $^{\mathrm{a}}$Dipartimento di Fisica Teorica, Universit\`a di Torino}\\
{\it and INFN, Sezione di Torino, 
Via P. Giuria 1, 10125 Torino, Italy}
\\
{\it $^{\mathrm{b}}$Dipartimento di Fisica, Universit\`a di Ferrara}\\
{\it and INFN, Sezione di Ferrara, I-44100 Ferrara, Italy}
\end{tabular}
}

\date{\today}
\maketitle
\begin{abstract}
We  discuss the properties of strange matter, in
particular the minimum of the energy per baryon number as a function of the 
strangeness fraction. We utilize both the MIT bag model and the Color 
Dielectric Model and compare the energy per baryon with the masses of 
hyperons having the 
corresponding strangeness fraction, which are coherently calculated within
both models. We also take into account the perturbative exchange 
of gluons. The results obtained in the two approaches allow to discuss 
the stability of strangelets. While the MIT bag model and the double minimum version of the Color Dielectric Model allow the existence of strangelets, the single minimum version of the Color Dielectric Model excludes this possibility.

\end{abstract}
\vspace{0.8cm}
\section{Introduction}

It has been suggested long ago~\cite{Greiner88} that hypermatter and/or 
strange quark matter can be realized in central relativistic heavy ion 
collisions, in 
which either multi--hypernuclear objects (strange hadronic matter) or
strangelets (strange multiquark droplets) could be produced. The occurrence
of the latter would be an unambiguous signature of the transient existence
of a deconfined, strangeness rich, state
of quark gluon plasma (QGP). Of course the relevance of these objects highly
depends on their stability after formation, and hence on the possibility of
their detection.

An enhancement of strange particle production has been observed since the 
early 90's in many experiments with relativistic heavy ion 
collisions~\cite{expvecchi}.
Indeed a large number of $s{\bar s}$ pairs can be produced in a single central
event: the antiquarks $\bar s$ combine with the abundantly available light 
quarks $u$ and $d$, thus producing antikaons, which rapidly leave the fireball
region. The residual system turns out then to be a strangeness rich matter 
which hadronizes with a copious production of strange particles, especially 
$K$ mesons and $\Lambda$ hyperons: strongly enhanced yields have been by now
assessed~\cite{expnuovi}.
Yet this outcome cannot be considered as a reliable signature of quark--gluon
plasma (QGP) formation, since kaons and hyperons can be produced in hadronic
reactions as well, before the nuclear fireball reaches 
equilibrium~\cite{Mattie89}.

The case would be different if, after formation of the deconfined plasma, this
strangeness rich matter could coalesce into colorless multiquark states, the 
so--called strangelets. The prompt anti-kaon (and also pion) emission from the 
surface of the fireball could, in addition, rapidly cool the QGP, thus 
favouring the condensation into metastable or stable droplets of strange 
quark matter~\cite{Greiner88}.

Long after the first definition of strangelets, by Bodmer~\cite{Bodmer71} 
and Chin and Kerman~\cite{Chin79}, Witten conjectured~\cite{Witten84} that 
strange quark matter can be absolutely stable (namely more stable than 
ordinary nuclei), a result obtained within the MIT bag model with parameter
$B\simeq 58$~MeV~fm$^{-3}$. The main reason underlying this 
situation is the lowering of the Fermi energy introduced by the new degree
of freedom (the $s$-quark), which in turn lowers the global minimum of the
energy per baryon ($E/A_B$) with respect to ordinary nuclear matter.

It is worth mentioning that strangelets can also be produced by coalescence
of hyperons~\cite{Dover93}: it consists of the fusion of a few hyperons and 
nucleons and does not imply the existence of the QGP phase: however this 
process can typically produce only small baryon numbers (e.g. the H dibaryon).
Hence the detection of heavy strangelets ($ A_B\ge 20-30$) should remain an 
unambiguous signature of QGP formation.
For a review on the properties of strange matter and strangelets we refer
the reader to the papers by Greiner and 
Schaffner~\cite{Greiner96,Greiner:1998jy}. 

Up to now the properties and stability of strangelets have been discussed
within the MIT bag model or, similarly, a Fermi gas model stabilized by the 
vacuum pressure $B$; the pioneering work by Fahri and Jaffe~\cite{Fahri84}
 also includes ${\mathcal{O}} (\alpha_s)$ corrections to the properties of bulk 
strange matter and finds the stability conditions: according to these authors 
 heavy, slightly positively charged, strangelets could be more stable than
ordinary nuclei. Greiner {\it et al.}~\cite{Greiner88b} find
(in pure MIT bag, without color exchange contributions) that also light 
strangelets (with $A_B\le 20$) are metastable due to finite size and shell 
effects. A detailed calculation of strangelet properties within the MIT bag
model, including shell effects and all the hadronic decay channels has been
performed by J. Schaffner {\it et al.}~\cite{Schaffner97}: a valley of 
stability clearly appears for $A_B= 5\div 16$ with charge fraction $Z/A$
between 0 and $-0.5$. On the other hand, strangelets having a larger mass should be positively charged according to the results of Ref.~\cite{Madsen00}. 

In this paper we want to discuss the properties of strange matter, in
particular the minimum of the energy per baryon number as a function of the 
strangeness fraction. We utilize both the MIT bag model and the Color 
Dielectric Model (CDM) and compare the equilibrium energy of the strange 
matter with the masses of hyperons having the same strangeness fraction; 
the latter are coherently calculated within both the MIT and CDM models.
 The main goal is to find out whether and to which extent 
the stability of strange matter and/or strangelets with respect to ordinary 
hadronic matter depends on the model employed to describe the confined 
system of quarks.

We consider homogeneous quark matter made up of $u$, $d$ and $s$ quarks. 
We assume that, during a high energy collision between heavy ions, this state 
of matter, if formed, can only survive for a very short time, so that it 
has no time to reach $\beta$ equilibrium; hence we do not impose chemical 
equilibrium on the density of the strange quarks, limiting 
 ourselves to assume that there exists in the system  a strange fraction 
$R_s=\rho_s/\rho$, $\rho$ being the total baryon density of quarks and 
$\rho_s$ the baryonic density of strange quarks. 
Our aim is to find out whether this system is more or 
less stable than hyperons, in order to understand which state is more likely 
to be produced in heavy ion collisions, either hyperons and strange mesons or
 strangelets. 

The energy per baryon number in the mean field 
approximation is a function of the baryon density, $\rho$, and of $R_s$. 
Thus we consider the surface which represents the energy per particle, 
$E/A_B$, versus $\rho$ and $R_s$ and concentrate on its 
sections with constant $R_s$: the resulting curves  represent 
the energy per baryon number, at a fixed $R_s$, as a function of $\rho$, and 
in general they present a minimum. The minimal energy per baryon number can 
then be studied as a function of the strange fraction $R_s$. 
In order to describe the system of three flavors  we shall employ
 two different models: the MIT bag model and the CDM. We also consider the 
effect induced by the introduction of perturbative gluons. Electromagnetic interaction has been neglected in this paper and therefore the minimum of the energy corresponds to an equal number of $u$ and $d$ quarks.
In the MIT bag model, $u$ and $d$ quarks are massless, while 
in the CDM they are massive: thus it will be interesting to compare the 
results obtained in each case, in order to find out the role of light quark 
masses on the bulk properties of strange quark matter. Another difference between the two models is that in the MIT bag model there is a sharp transition from the inner region to the outer region of the nucleon. In the CDM, on the contrary, both a two phase and a single phase scenario can be obtained. As we will see, these two possibilities give rather different results for the stability of strangelets.

Since we consider an infinite and homogeneous system, while 
strangelets are finite objects, one should be careful in drawing conclusions
about the stability of strangelets on the basis of strange matter 
stability; however the energy of the infinite system appears to be a lower
limit with respect to the envelop of  strangelet energies versus 
strangeness fraction: the latter was nicely illustrated by Schaffner {\it et 
al.}~\cite{Schaffner97} calculating the strangelet masses within the MIT bag 
model with shell mode filling. We report their result in Fig.~\ref{fig1}, 
which will
be useful for further comparison with our results. Here we simply recall that
 surface effects, which we do not consider, would increase the energy curves
of bulk matter, typically of 50-100~MeV: hence, if hyperons should 
turn out to be more stable than strange matter, then this would exclude also
the stability of strangelets. If, on the contrary, strange matter is more 
stable, then this provides only an indication in favour of stable strangelets.

\section{Strangelets in the MIT bag model}
The MIT bag model has been widely utilized in the past, both for strange 
matter and for strangelets. For detailed derivations we refer the reader 
to the literature~\cite{Witten84,Fahri84,Schaffner97,DeGrand75}.

\subsection{MIT bag model without gluons}

We shall use here the simplest version of the MIT bag model, not including
one gluon exchange corrections. Therefore, we have two model parameters: 
the vacuum pressure $B$ and the strange quark mass $m_s$. We have used 
different values of these parameters, in order to discuss various possible scenarios.

Calculations of hadron spectroscopy~\cite{DeGrand75} indicate for $B$ a value of the 
order of 60 MeV/fm$^3$. However this value is generally not adopted in 
quark matter calculations, because it produces  too much binding (notice 
that it is close to the 58 MeV/fm$^3$ of Witten's conjecture);
 calculations of the hadronic structure functions~\cite{Thomas95} suggest  a value of 
$B\simeq 100$~MeV/fm$^3$. Hence, besides $B=60$~MeV/fm$^3$, we consider also 
$B=100$~MeV/fm$^3$ and an additional value, $B=150$~MeV/fm$^3$. 
The latter  does not correspond to calculations of physical 
quantities, but it has been indicated in the literature~\cite{Satz82} 
as a sensible value in a comparison between the results of MIT bag model and 
lattice QCD at finite temperature.  

The single flavor contribution to the energy density of the system is given by:

\beq
\epsilon_f=6\int\frac{d\vec k}{\left(2\pi\right)^3}E_f\left(k\right)\theta\left(k_{F_f}-k\right),
\eeq
where $E_f(k)=\sqrt{k^2+m_f^2}$ and $k_{F_f}$ is the Fermi momentum of flavor $f$.
We use natural units ($\hbar=c=1$).
Since $u$ and $d$ quarks are massless, their contribution to the energy 
density is:
\beq
\epsilon_{u,d}= \frac{3}{\left(2\pi\right)^2}\,k_{F_{u,d}}^4\,,
\eeq
while the  strange quark contribution reads:
\beq
\epsilon_s=\frac{3}{8\pi^2}
\left[m_s^4\ln\left(\frac{m_s}{k_{F_s}+\sqrt{k_{F_s}^2+m_s^2}}\right)
+k_{F_s}\sqrt{k_{F_s}^2+m_s^2}\left(2k_{F_s}^2+m_s^2\right)\right]\,,
\eeq
 $m_s$ being the mass
of the quark $s$. 

The total energy density of our system turns then out to be:
\beq
\epsilon_{tot}=2\epsilon_{u,d}+\epsilon_s+B \,.
\eeq

The dependence of the above formula on $R_s$ and $\rho$ can be easily found 
by recalling that:
\beq
\rho_s=R_s\rho
\eeq
and:
\bea
k_{F_s}&=&\left(3\pi^2\rho_s\right)^{1/3}\\
k_{F_{u,d}}&=&\left(\frac{3\pi^2}{2}\rho\left(1-R_s\right)\right)^{1/3},
\eea
$\rho$ being the total baryon number density in the system ($\rho=A_B/V$).
In the above the color degeneracy and baryon number $1/3$ of the quarks 
have been taken into account.

From the above formulas we calculate the energy per baryon number to be:
\beq
\frac{E_{tot}}{A_B}=\frac{\epsilon_{tot}}{\rho}.
\label{Eperpart}
\eeq
In Fig.~\ref{fig2} the results of the minimal energy per baryon 
(\ref{Eperpart}) corresponding to $B=60, 100, 150$~MeV/fm$^3$ are shown as a
function of $R_s$.
For each value of $B$ we explore three different values of the strange 
mass, $m_s=100, 200, 300$~MeV, and we compare these results with the 
experimental nucleon and hyperon masses (full circles).
We have also evaluated, according to formula (3.6) of Ref.~\cite{DeGrand75},
the baryonic masses which are obtained within the same model employed for
bulk strange matter, using the same sets of bag parameter and strange quark mass.
Here we set to zero the perturbative one gluon exchange corrections, which are, instead, taken into
account by De Grand {\it et al.}~\cite{DeGrand75} and which will be 
considered in the next subsection.
For baryons, the fraction $R_s$ is assumed to correspond to the fraction of 
$s$ quarks with respect to $u$ and $d$ quarks in the considered hadron.

%

As it appears from the figure, the three lines corresponding to the 
different values of $m_s$ are much lower than the experimental hyperon masses 
for $B=60$~MeV/fm$^3$ and $B=100$~MeV/fm$^3$, while this is not the case for 
$B=150$~MeV/fm$^3$ and $m_s$=300 MeV. In this instance the experimental hyperon masses are 
lower than the bulk matter energy. However, if we compare the energy of 
strange matter with the corresponding theoretical masses of the various 
hyperons, we find that strange matter is {\it always} lower in energy, and
thus more stable. In agreement with the results of Ref.~\cite{Schaffner97}, which are shown in Fig.~\ref{fig1}, we can conclude that the MIT bag model without perturbative gluon corrections allows the existence of strangelets. It might be interesting to notice also that in the MIT bag model a minimum at a finite 
value of $R_s$ is always present.
The only parameter set not showing a minimum corresponds to $B=60$~MeV/fm$^3$ 
and $m_s=300$~MeV.

\subsection{MIT bag model with perturbative gluons.}

We consider now the effects of the introduction of perturbative 
gluons  in the calculation. At first order in $\alpha_s$, two contributions to the energy can be considered, the direct and the exchange one. Since the system is globally colorless the direct term vanishes, while the exchange one gives the following contribution to the energy density of quarks of flavor $f$ ~\cite{Fahri84}:\footnote{
Actually the third term in square brackets has a different sign, with 
respect to Ref.~\cite{Fahri84}, in agreement with a recent re-derivation
of the formula~\cite{Andrea}.}
\bea
\epsilon_f^{OGE}&&=-\frac{\alpha_s}{\pi^3}m_f^4\left\{x_f^4-\frac{3}{2}
\left[\ln\left(\frac{x_f+\eta_f}{\eta_f}\right)-x_f\eta_f\right]^2+\right.
\nonumber\\
&&\qquad \left. +
\frac{3}{2}\ln^2\left(\frac{1}{\eta_f}\right)
-3\ln\left(\frac{\mu}{m_f\eta_f}\right)\left[\eta_fx_f
-\ln\left(x_f+\eta_f\right)\right]\right\}\,.
\label{ogemit}
\eea
Here:
\bea
x_f &=& \frac{k_{F_f}}{m_f}
\nonumber\\
\eta_f &=& \sqrt{1+x_f^2}.
\nonumber
\eea
and $\mu$ is a renormalization scale, for  which we choose the value
$\mu=313$~MeV, according to Ref.~\cite{Fahri84}. 
In Fig.~\ref{fig3} we show $\epsilon^{OGE}$ as a function of $k_F$ for various values of $m$. It is evident that, for small values of $m$, the contribution is always repulsive, while for $m\gtrsim200$ MeV there exists a range of densities for which it is attractive. Hence the effect of including perturbative gluons in the energy density of the system will crucially depend on the fraction of strange quarks, as well as on the value of $m_s$.
If the quark mass vanishes ($m_f=0$) eq.~(\ref{ogemit})
reduces to:
\beq
\epsilon_{u,d}^{OGE}= \frac{\alpha_s}{\pi^3}k_{F_{u,d}}^4.
\eeq
Yet in the following we shall use a small nonzero mass for the $u$ and $d$ quarks, $m_{u,d}=4$~MeV.
For sake of illustration, we adopted two different values for $\alpha_s$, 
a small perturbative value ($\alpha_s=0.5$) and the canonical value which 
was employed  by DeGrand {\it et al.}~\cite{DeGrand75} ($\alpha_s=2.2$), 
to reproduce the hyperon masses.
The corresponding results are illustrated in Figs.~\ref{fig4} and \ref{fig5}.
 
By comparing Fig.~\ref{fig2} and \ref{fig4} 
we can see that the gluon effect is always repulsive at low $R_s$, while at large $R_s$ it is repulsive for $m_s=100,200$~MeV, and attractive for $m_s=300$~MeV. This is even more evident in Fig.~\ref{fig5}, 
where the effect of gluons is stronger, due to the larger value of $\alpha_s$: 
in this case the attractive effect at large $R_s$ is so 
important that the curve corresponding to $m_s=300$~MeV becomes the lowest 
one, in contrast with the situation shown in Figs.~\ref{fig2} and \ref{fig4}.

From Fig.~\ref{fig4} we can see that, even after the inclusion of 
perturbative gluons, strangelets are more stable than hyperons
for almost all values of the model parameters. However, when we 
use the stronger coupling of Fig.~\ref{fig5} the stability of strange
matter (and hence strangelets) becomes questionable, particularly 
for low values of the strange mass $m_s$. Only for $m_s=300$~MeV the
theoretical masses of hyperons always lie above the energy of bulk matter
(not so the experimental masses). It is worth keeping in mind that among 
all the situations illustrated here, only the upper left panel of 
Fig.~\ref{fig5} utilizes parameters close to the ones of 
DeGrand {\it et al.}~\cite{DeGrand75} ($B=59$~MeV/fm$^3$, $m_s=279$~MeV, 
$\alpha_s=2.2$): indeed the stars ($m_s=300$~MeV) reproduce fairly well 
the experimental masses, and for this choice of parameter values the 
bulk strange matter turns out to be favoured.

From this analysis we can conclude (in agreement with previous findings)
that, apart from rather extreme choices of the model parameters, metastable strangelets can exist in the MIT bag model.

\section{Strangelets in the Color Dielectric Model}

The Color Dielectric Model provides absolute confinement of quarks through
their interaction with a scalar field $\chi$ which represents a 
multi--gluon state and produces a density dependent constituent mass 
(see for example the review articles~\cite{Wilets}--\cite{Pirner92}).
Several versions of CDM have been employed  to calculate hadronic 
properties~\cite{Chanfray84}--\cite{Barone94}
and quark matter equation of state~\cite{Drago95}--\cite{Drago01}.

The typical Lagrangian of the CDM is the following~\cite{Aoki91,Barone95}:
\bea
{\cal L} &&= \sum_{f=u,d,s}{\bar{\psi_f}} i\gamma^{\mu}\left(\partial_{\mu}
-ig_s\frac{\lambda^a}{2}A^a_{\mu}\right)\psi_f
-\frac{g f_{\pi}}{\chi}\sum_{f=u,d}{\bar{\psi_f}} \psi_f
-m_s\left(\chi\right) {\bar{\psi_s}} \psi_s +
\nonumber\\
&&\quad
+\frac{1}{2}\left(\partial_{\mu}\chi\right)^2-U\left(\chi\right)
-\frac{1}{4}\kappa\left(\chi\right)F^a_{\mu\nu}F^{a \mu\nu}\,,
\label{CDMlagr}
\eea
where $\psi_f$ are the quark fields, $A^a_{\mu}$ is the (effective) gluon 
field, $F^a_{\mu\nu}$ its strenght tensor and $\chi$ is the color dielectric 
field. 

The $u$ and $d$ quark masses are the result of their interaction with the 
$\chi$--field and read:
\beq
m_{u,d}=\frac{g f_{\pi}}{\chi},
\label{udmass}
\eeq
where $g$ is a parameter of the model and $f_\pi$ the pion decay constant, 
which is fixed to its experimental value, $f_\pi=93$~MeV.\footnote{
In some chiral invariant versions of the CDM the mass term is also coupled
to the usual $(\sigma,\vec\pi)$ fields~\cite{Neuber93,Barone94,Drago01}.}
 For the strange quark mass we consider two different 
versions of the 3--flavors CDM, namely a {\it scaling model}, with
\beq
 m_s=\frac{g f_{\pi}+\Delta m}{\chi}\equiv \frac{g' f_{\pi}}{\chi}\,,
\label{smass1}
\eeq
and a {\it non--scaling model}, with a constant shift of the $s$--mass
with respect to the $u,d$--one:
\beq
 m_s=\frac{g f_{\pi}}{\chi} +\Delta m \equiv m_{u,d}+\Delta m \,.
\label{smass2}
\eeq
In the above $g'$ and/or $\Delta m$ is another parameter of the model.

Concerning the color dielectric field, there exist several versions, both 
for its coupling to the gluon tensor and for the potential $U(\chi)$. 
We adopt here both the single minimum (SM), quadratic potential:
\beq
U_{SM}\left(\chi\right)=\frac{1}{2}M^2\chi^2\,,
\label{sm}
\eeq
which introduces the third parameter of the model, $M$ (the mass of the 
glueball), and the double minimum (DM), quartic potential:
\beq
U_{DM}\left(\chi\right)=\left(\frac{1}{2}\frac{M^2}{\chi_0^2}-\frac{3B}{\chi_0^4}\right)\chi^4+\left(\frac{4B}{\chi_0^3}-\frac{M^2}{\chi_0}\right)\chi^3+\frac{1}{2}M^2\chi^2.
\label{dm}
\eeq
The latter introduces an extra parameter, the bag pressure B, while the parameter $\chi_0$ is used to make the ratio $\chi/\chi_0$ dimensionless.
 The color--dielectric function, $\kappa(\chi)$, is usually 
assumed to be a quadratic or quartic function of $\chi$: we will use both
options and hence we set:
\beq
\kappa\left(\chi\right)=\left(\frac{\chi}{\chi_0}\right)^{\beta},
\qquad\qquad{\mathrm{ with}}\qquad \beta=2,4\ .
\label{cappa}
\eeq

The field equations will be solved in the mean field 
approximation and without gluons: the latter are subsequently taken into 
account as a perturbation. For homogeneous quark matter the color 
dielectric field must obey the equation:
\beq
\left.\frac{dU\left(\chi\right)}{d\chi}\right|_{\chi={\bar\chi}} =
\frac{gf_{\pi}}{\bar{\chi}^2}\sum_{f=u,d}\langle\bar{\psi_f}\psi_f\rangle
+ \frac{g (g')f_{\pi}}{\bar{\chi}^2}\langle\bar{\psi_s}\psi_s\rangle\,,
\label{meanfield}
\eeq
where the second term on the r.h.s. will contain $g$ or $g'$, depending on the
choice of the non--scaling model ($g$) or of the scaling one ($g'$), 
respectively. In the above equation:
\beq
\langle\bar{\psi_f}\psi_f\rangle\equiv 
\rho_S^f\left({\bar\chi}\right)= 6\int\frac{d\vec k}{(2\pi)^3}
\frac{m_f(\bar{\chi})}{\sqrt{{\vec k}^2+m_f^2(\bar{\chi})}}
\theta\left(k_{F_f}-k\right)
\label{scalden}
\eeq
is the scalar density of quarks $f$.

The unperturbed (i.e. without gluon contribution) energy density reads:
\bea
\epsilon_0=&&\sum_{f=u,d,s}\frac{3}{8\pi^2}\left\{ m_f^4\ln\left(\frac{m_f}
{k_{F_f}+\sqrt{k_{F_f}^2+m_f^2}}\right)\right.
\nonumber\\
&& \left.
+ k_{F_f}\sqrt{k_{F_f}^2+m_f^2}\left(2k_{F_f}^2+m_f^2\right)\right\}
+U\left(\bar{\chi}\right)\,,
\label{eps0f}
\eea
the quark masses being given by eqs.~(\ref{udmass}) and (\ref{smass1}) 
[or (\ref{smass2})] with $\chi=\bar{\chi}$.

Beyond $\epsilon_0$ we have perturbatively taken 
into account, to order $\alpha_s$, the exchange of gluons, whose contribution
to the energy density of an infinite, color singlet system is, in analogy with eq.~(\ref{ogemit}):
\bea
\epsilon_{OGE}&=&\frac{E_{OGE}}{V} = 
\sum_{f=u,d,s} 6\int \frac{d\vec q}{(2\pi)^3} V_{OGE}(\vec q)
\theta\left(k_{F_f}-k\right)
\label{epsoge1}\\
&& = -\frac{\widetilde\alpha_s}{\pi^3}\sum_{f=u,d,s} m_f^4\left\{
x_f^4 -\frac{3}{2}\left[ \ln\left(\frac{x_f+\eta_f}{\eta_f}\right) 
-x_f\eta_f\right]^2 + \right.
\nonumber\\
&& \left. + \frac{3}{2}\ln^2\left(\frac{1}{\eta_f}\right) -
3\ln\left(\frac{\mu}{m_f\eta_f}\right)\left[ x_f\eta_f -
\ln\left(x_f+\eta_f\right)\right] \right\}\,.
\label{epsoge2}
\eea

Here the notations are the same as in eq.~(\ref{ogemit}), but for the CDM
definition of the constituent quark masses (in MFA); the effective strong
coupling constant, according to the CDM Lagrangian (\ref{CDMlagr}), reads:
\beq
\widetilde\alpha_s = \alpha_s\left(\frac{\chi_0}{\bar{\chi}}\right)^{\beta}
\label{alfaeff1}
\eeq
and becomes very large at small densities, where $\bar{\chi}\to 0$. 
As already remarked above, eq.~(\ref{epsoge1}) only 
contains the exchange term of OGE, the direct one vanishing for infinite 
quark matter: at small baryonic densities the attractive electric 
contribution dominates the energy density, while the repulsive magnetic 
contribution becomes the dominant one at large densities.

Indeed the divergent behavior of the electric term  for $\rho\to 0$ could
prevent the reliability of a perturbative treatment of OGE. We have 
overcome this difficulty by taking into account the Debye screening of the 
gluon propagator in the presence of a polarized medium. This can be 
achieved by replacing (\ref{alfaeff1}) with the new effective coupling:
\beq
\label{alfaeff2}
\alpha_s^{eff}(q) = \widetilde\alpha_s\frac{q^2}
{q^2+\frac{1}{2}\sum_{f=u,d,s} 16\widetilde\alpha_s m_f k_{F_f}^2
g(q/k_{F_f})}\,,
\eeq
$g(y)$ being the static limit of the polarization propagator~\cite{FetWal}:
\beq
g\left(y\right)=\frac{1}{2}-\frac{1}{2y}\left(1-\frac{1}{4}y^2\right)
\ln\left|\frac{1-\frac{1}{2}y}{1+\frac{1}{2}y}\right|\,.
\label{gpol}
\eeq

Actually this expression should be utilized in $V_{OGE}(\vec q)$ of 
eq.~(\ref{epsoge1}) and again integrated. For simplicity, since the $q$--integration is extended only up 
to $k_{F_f}$ and the function $g(y)$ varies at most by $9\%$ in the range
$0\le y \le 1$, we have adopted $q=k_{F_f}$.\\
The flavor dependent effective coupling reads therefore:
\beq
\label{alfaeff3}
\alpha_{s,f}^{eff} = \alpha_s\left(\frac{\chi_0}{\bar{\chi}}\right)^{\beta}
\frac{k_{F_f}^2}
{k_{F_f}^2+\frac{1}{2}\sum_{f=u,d,s} 16\alpha_s
\left(\chi_0/\bar{\chi}\right)^{\beta} m_f k_{F_f}^2 g(1)}\,.
\eeq
We notice that, taking into account the dependence of $\bar{\chi}$
upon the density, this new effective coupling vanishes at small 
densities: $\alpha_{s,f}^{eff}\sim\rho^{\frac{1}{3}}$, while it goes to zero as $k_{F_f}^{-\beta}$ at high densities. At variance with the Debye screening in electrodynamics, which is mainly relevant at large densities, in the CDM the screening is large at small densities too, due to the divergence of $\widetilde\alpha_s$, which enhances the effect of medium polarization even at small densities.
With these ingredients we shall compare our results  for the minimal energy
per baryon number with the hyperon masses, as they have been evaluated in 
the CDM. There exist in the literature two distinct calculations of 
this type and we shall consider both cases.

\subsection{Stability of strangelets in the CDM: I}

We consider here the work by Aoki {\it et al.}~\cite{Aoki91}: these authors
solve self-consistently the mean field equations for quarks, color dielectric field and gluons, starting from a CDM Lagrangian with the Double Minimum potential (\ref{dm})
for the color dielectric field. Concerning the color dielectric function, Aoki {\it et al.} choose $\beta=2$ in eq.~(\ref{cappa}).

In order to reproduce the masses of the octet and decuplet baryons (in
particular hyperons) the authors of Ref.~\cite{Aoki91} employ both the scaling
model and the non--scaling one. In addition, two different sets of 
parameters are used, whose values are dictated by two different and extreme choices 
for the ``bag'' parameter $B$: 
$B^{1/4}=0$~MeV, with two degenerate vacua, and a large bag pressure, 
$B^{1/4}=103.5$~MeV. The latter value of B is chosen to be as large as possible, but with the requirement that the two-phase picture must hold inside hadrons. In their calculation only the strange quark mass has to
be considered as a truly free parameter, the remaining ones having been fixed 
in a previous work on the non--strange baryons~\cite{Aoki90}.
We only perform calculations with the second set of parameters, corresponding to the non-zero value of the bag pressure, since $B=0$ does not give quark confinement at low densities.\\
The field equation for the color dielectric field, eq. ({\ref{meanfield}), now becomes:

\beq
4\left(\frac{1}{2}\frac{M^2}{\chi_0^2}-\frac{3B}{\chi_0^4}\right)\bar\chi^3+3\left(\frac{4B}{\chi_0^3}-\frac{M^2}{\chi_0}\right)\bar\chi^2+M^2\bar\chi=\frac{gf_{\pi}}{{\bar{\chi}}^2}
\sum_{f=u,d,s} \rho_S^f\left({\bar\chi}\right)
\label{nonscalmfdm}
\eeq
for the non-scaling model, and:

\beq
4\left(\frac{1}{2}\frac{M^2}{\chi_0^2}-\frac{3B}{\chi_0^4}\right)\bar\chi^3+3\left(\frac{4B}{\chi_0^3}-\frac{M^2}{\chi_0}\right)\bar\chi^2+M^2\bar\chi=\frac{gf_{\pi}}{{\bar{\chi}}^2}\left[
\rho_S^u\left({\bar\chi}\right)+\rho_S^d\left({\bar\chi}\right)\right] +
\frac{g'f_{\pi}}{{\bar{\chi}}^2}\rho_S^s\left({\bar\chi}\right)
\label{scalmfdm}
\eeq
for the scaling model, respectively.
The parameters of Ref.~\cite{Aoki90} (with $B^\frac{1}{4}=103.5$~MeV) appropriately converted to the notation of the 
present work, are the following ($\alpha_s=g_s^2/4\pi$):\\
\bea
g=43.7\,\mathrm{MeV}\qquad M=1177\,\mathrm{MeV}
\qquad \chi_0=47.1\,\mathrm{MeV}\qquad\alpha_s=2.0
\nonumber
\eea

Hence, we consider cases B and D of the work of Aoki {\it et al.}:
\begin{description}

\item{B:}
scaling model, $g'=138.9$ MeV;
\item{D:}
non--scaling model, $\Delta m=212$~MeV.
\end{description}

We report in Table I the masses of a few hyperons (the ones of interest for 
the present work) as obtained in Ref.~\cite{Aoki91}, together with their 
experimental values: the agreement is remarkable.

\begin{table}[t]
\begin{center}

\begin{tabular}{c|cccc}
\hline
\hline
Hyperon & N &$\Lambda$ &$\Xi$ &$\Omega$ \\
\hline
\hline \\
$R_s$ &0 &0.33 &0.67 &1 \\
\hline
\hline \\
$m_{exp}$~(MeV) & 938 &1115 &1314 &1672 \\
\hline
\hline \\
$m_{B}$~(MeV) & 938 & 1161 & 1346 & 1639 \\
\hline
\hline \\
$m_{D}$~(MeV) & 938 & 1113 & 1307 & 1671 \\
\hline
\hline
\end{tabular}
\end{center}
\begin{center}
\footnotesize
Table I. Strangeness fraction, experimental masses and theoretical masses of
hyperons, according to the calculation of Ref.~\cite{Aoki91}.
\end{center}
\end{table}

We evaluate the minimum energy per baryon number using cases B and D, both without and with the perturbative
correction of the effective gluon exchange.\footnote{In the Appendix we discuss the existence of multiple solutions for field equations when the potential $U(\chi)$ has a double minimum.} In Fig.~\ref{fig6} we compare our curves with 
the masses calculated in Ref.~\cite{Aoki91}, which always include, as already 
mentioned, the gluon field correction in a self--consistent approach.
In addition to the theoretical masses of hyperons, we also show in the figure 
their experimental values (represented by full circles) which, according to 
Table~I, are always rather  close to the calculated ones.

 As we can see from Fig.~\ref{fig6}, the effect of perturbative gluons 
in this model is very small, due to the Debye screening, which we have included. 
 Whether or not we 
take into account gluon corrections, strange matter always appears to be more
stable than baryons. 

Hence from the cases considered here we can conclude that this version of the CDM seems to 
favour strangelets as a metastable form of matter. This is due to the fact that when the DM potential is used to study hadrons, i.e. confined objects, a large contribution to the hadronic mass is given by the space fluctuations of the fields. When this version of the model is used to describe infinite quark matter, these contributions vanish due to the homogeneity of the system. For this reason, deconfined matter is favoured in this version of the model. This finding is consistent with the results of Refs. ~\cite{Drago95,Barone95}. It is also worth mentioning that, as shown in Fig.~\ref{fig6}, the minimum of the energy per baryon number corresponds to $R_s=0$, at variance with the results of the MIT bag model.

\subsection{Stability of strangelets in the CDM: II}

In this section we follow the approach of 
J.~McGovern~\cite{McGovern}. The model Lagrangian is still the one reported 
in eq.~(\ref{CDMlagr}) with $\beta=4$ in the color dielectric function.
McGovern uses only the scaling model and the Single Minimum potential, 
with different values of the parameters. In this case 
the behavior of $\alpha_s^{eff}(\bar\chi)$ is even more divergent, for small 
densities, than in the case $\beta=2$ previously discussed: hence the use of Debye
screening in the effective strong coupling constant is mandatory.

In Ref.~\cite{McGovern} two different sets for the model parameters are used:
they allow to satisfactorily reproduce the splittings between hyperon masses, 
but the absolute values of the masses themselves are generally too large. The 
latter are reported in Table II.\\
In this model, the field equation for the scalar field becomes:
\beq
M^2{\bar{\chi}}=\frac{gf_{\pi}}{{\bar{\chi}}^2}\left[
\rho_S^u\left({\bar\chi}\right)+\rho_S^d\left({\bar\chi}\right)\right] +
\frac{g'f_{\pi}}{{\bar{\chi}}^2}\rho_S^s\left({\bar\chi}\right).
\label{scalmfsm}
\eeq

\begin{table}[t]
\begin{center}

\begin{tabular}{c|cccc}
\hline
\hline
Hyperon & N &$\Lambda$ &$\Xi$ &$\Omega$ \\
\hline
\hline\\
$R_s$ &0 &0.33 &0.67 &1 \\
\hline
\hline\\
$m_{exp}$~(MeV) & 938 &1115 &1314 &1672 \\
\hline
\hline\\
$m_{\mathrm{set~I}}$~(MeV) & 1227 & 1427 & 1638 &1866 \\
\hline
\hline\\
$m_{\mathrm{set~II}}$~(MeV) & 1268 & 1417 & 1691 & 1928\\
\hline
\hline

\end{tabular}
\end{center}
\begin{center}
\footnotesize
Table II. Strangeness fraction, experimental masses and theoretical masses 
of hyperons, as calculated in Ref.~\cite{McGovern}.
\end{center}
\end{table}

The parameter sets used in Ref.~\cite{McGovern} are:\\

Parameter set I:
\bea
g=46.7~\mathrm{MeV}\qquad M=2354~\mathrm{MeV}\qquad 
\chi_0=20~\mathrm{MeV}\qquad \alpha_s=0.3380
\nonumber
\eea
\\

Parameter set II:
\bea
g=107.527~\mathrm{MeV}\qquad M=1000~\mathrm{MeV}
\qquad \chi_0=45.4~\mathrm{MeV} \qquad \alpha_s=0.3533
\nonumber
\eea 
We fix the ratio: $g'/g=1.58$, in agreement with Ref.~\cite{McGovern}.

As in the previous section, in Fig.~\ref{fig9} we compare our results for the minimum energy 
per baryon number both with the experimental and the theoretical masses.
As we can notice, also in this case the inclusion of perturbative gluons practically does not alter the quark matter energy, since here the Debye screening is overwhelming.  

Strange matter turns out to be well above the 
experimental hyperon masses and below the theoretical ones. We must remember that in our calculations we don't consider surface effects, which would increase the energy density of a finite system of $50\div100$ MeV, as we can deduce by comparing our results in the MIT bag model with Fig.~\ref{fig1}. In this perspective, only for $R_s\simeq\frac{2}{3}$ strangelets are (marginally) allowed by the present calculation and a more refined one, taking into account surface energy contributions, could clarify the situation.

We can therefore conclude that, but for $R_s\simeq\frac{2}{3}$, the existence of metastable strangelets is excluded in the Single Minimum version of the CDM. 

\section{Conclusions}
In this work we have computed the energy per baryon number of infinite quark matter having a fixed strangeness content. We have compared this quantity to the mass of hyperons having the same strangeness fraction and calculated within the same model and for the same parameter values that we adopt in our calculations.

Our analysis shows that the existence of strangelets
is supported only by those models which entail a two--phase picture of 
hadrons, namely which mantain a false vacuum inside hadrons.
This happens both in the MIT bag model, and in the Double Minimum version
of the Color Dielectric Model: as we have seen in Sections 2 and 3.1,
in these cases the minimum energy per baryon number versus the strangeness
fraction $R_s$ turns out to be lower than the masses of hyperons with the same
$R_s$.

As we have seen in Figs.~\ref{fig2},~\ref{fig4},~\ref{fig5},~\ref{fig6}, the mass gap between hyperons and strangelets can be as large as 300 MeV, and the metastability of strangelets would be confirmed even by taking into account surface energy effects, typically of the order of 100 MeV, in agreement with the findings of Ref.~\cite{Schaffner97}.

On the contrary, the Single Minimum version of the CDM
 does not support the existence of strangelets: 
independently of the set 
of parameter values, the computed masses of hyperons are larger than the corresponding strange matter energy by about $50 \div 100$ MeV. Only for $R_s=\frac{2}{3}$ the gap is 150 MeV and we cannot exclude completely the presence of strangelets with this strangeness fraction.

This outcome points out that the stability of strangelets appears to depend 
rather crucially on the model employed:
 CDM 
 supports stable strangelets only in the DM version, but not in the SM one.
This fact could set serious challenges to the search for strangelets in
relativistic heavy ion collisions, as a signature for the quark gluon plasma 
phase, out of which strangelets could be formed.

{\bf Acknowledgements}\\
It is a great pleasure to thank J\"urgen Schaffner-Bielich for many stimulating discussions.

\section{Appendix}
There can be multiple solutions of field equations, 
when for the dielectric field $\chi$ a potential
having two minima is used. This possibility 
is known since the first calculations within the
color-dielectric model~\cite{Dodd87,Bro}. 

 In the above quoted references, the solutions of the
field equations were non-topological solitons. In the present work
we are discussing plane-wave solutions for the quarks, which are considerably
simpler. The solutions of the field equations for $\chi$ are graphically shown in Fig.~\ref{fig10}, where they appear as the intersection of the dotted line with the dashed (at lower density) or continuous lines. As it appears from Fig.~\ref{fig10}, when the density is low
enough there can be three solutions. 
The solutions are characterized by their behavior at low density.
One solution corresponds to $\chi \rightarrow 0$ as $\rho \rightarrow 0$.
The other two solutions correspond to $\chi$ close to the
value of relative maximum of the potential
or to the relative minimum, respectively.
The solution in which $\chi$ is near the relative maximum is unstable,
and it corresponds to the unstable solitonic solution pointed
out in Ref.~\cite{Bro}. 

At low density there are therefore two stable solutions. The ``true''
solution is clearly the one of minimal energy. As it can be seen
in Fig.~\ref{fig11}, but for very small value of the density, the solution of minimal
energy is the one in which $\chi$ is close to the value of the relative 
minimum of the potential.

\begin{figure}
\begin{center}
\mbox{\epsfig{file=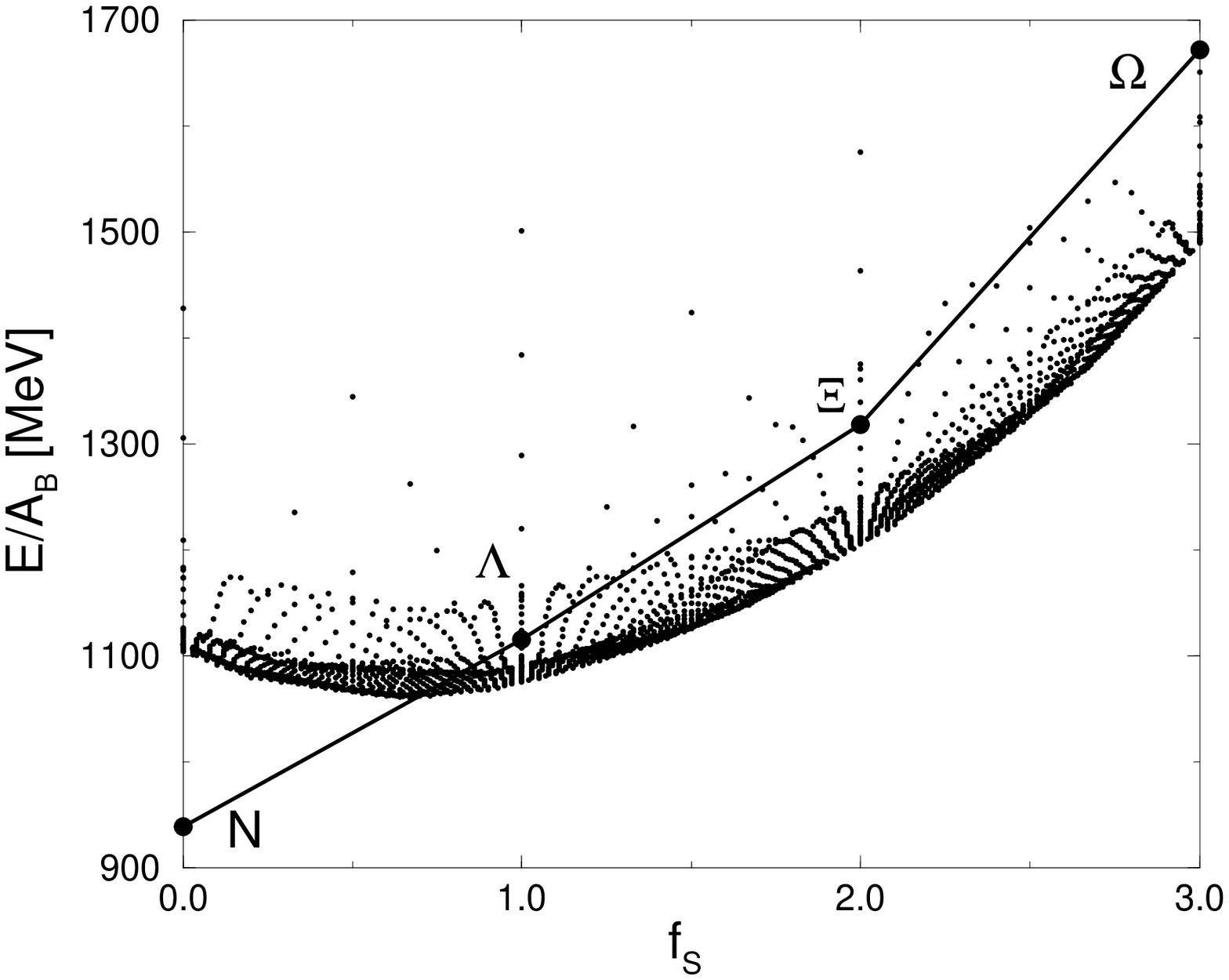,width=\textwidth}}
\end{center}
\vskip -0.8cm
\caption{The energy per baryon of all possible strangelets with $A_B< 40$,
for a bag constant $B^{1/4}=170$~MeV versus the strangeness fraction $f_s=3R_s=|S|/A_B$.
The solid line connects the masses of the nucleon and of the first lowest 
hyperons: it represents free baryonic matter. (Taken from 
Ref.~\cite{Schaffner97})
}
\label{fig1}
\end{figure}

\begin{figure}[htb]
\begin{center}
\mbox{\epsfig{file=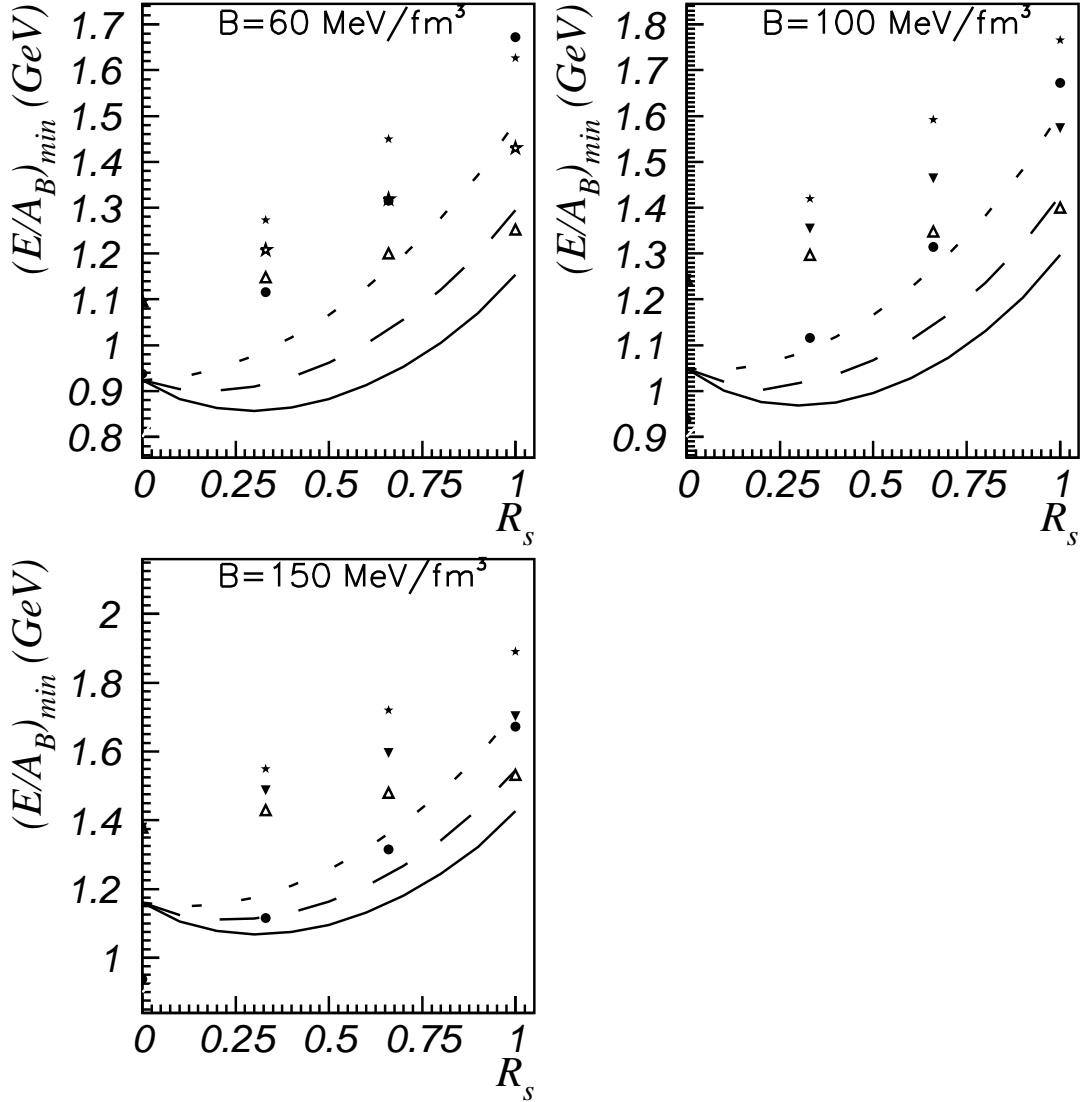,width=1\textwidth}}
\end{center}
\vskip -0.8cm
\caption{Minimal energy per baryon number in the MIT bag model, 
as a function of the strangeness
 fraction $R_s=\rho_s/\rho$, for various choices of the values of the 
model parameters. The continuous line corresponds to $m_s=100$~MeV, the 
dashed line to $m_s=200$~MeV and the dotted line to $m_s=300$~MeV. Full circles correspond to experimental masses, the other points to the masses 
evaluated in the model, with $m_s=100$~MeV (open triangles),
$m_s=200$~MeV (full triangles), $m_s=300$~MeV (stars), respectively.
}
\label{fig2}
\end{figure}
\begin{figure}[htb]
\begin{center}
\mbox{\epsfig{file=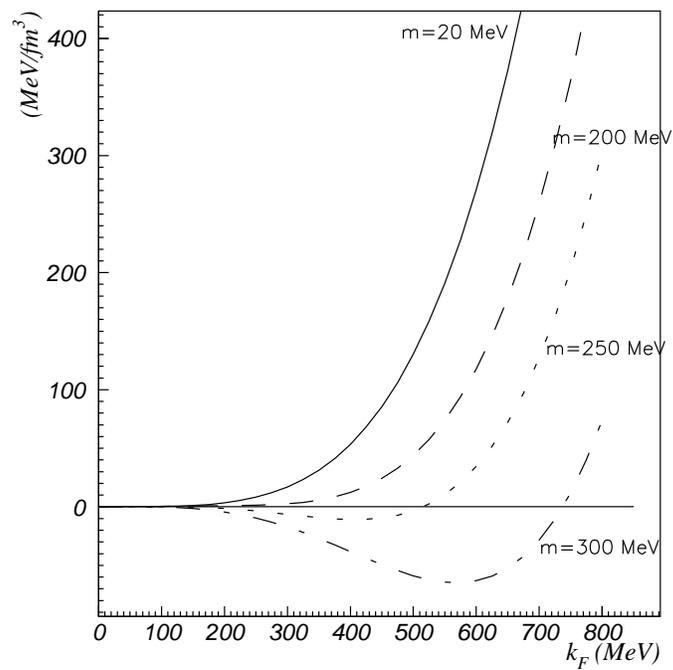,width=.6\textwidth}}
\end{center}
\vskip -.8cm
\caption{OGE contribution to the energy density divided by $\alpha_s$, $\epsilon^{OGE}/\alpha_s$, as a function of $k_F$ for different values of the quark mass.}
\label{fig3}
\end{figure}
%
\begin{figure}[htb]
\begin{center}
\mbox{\epsfig{file=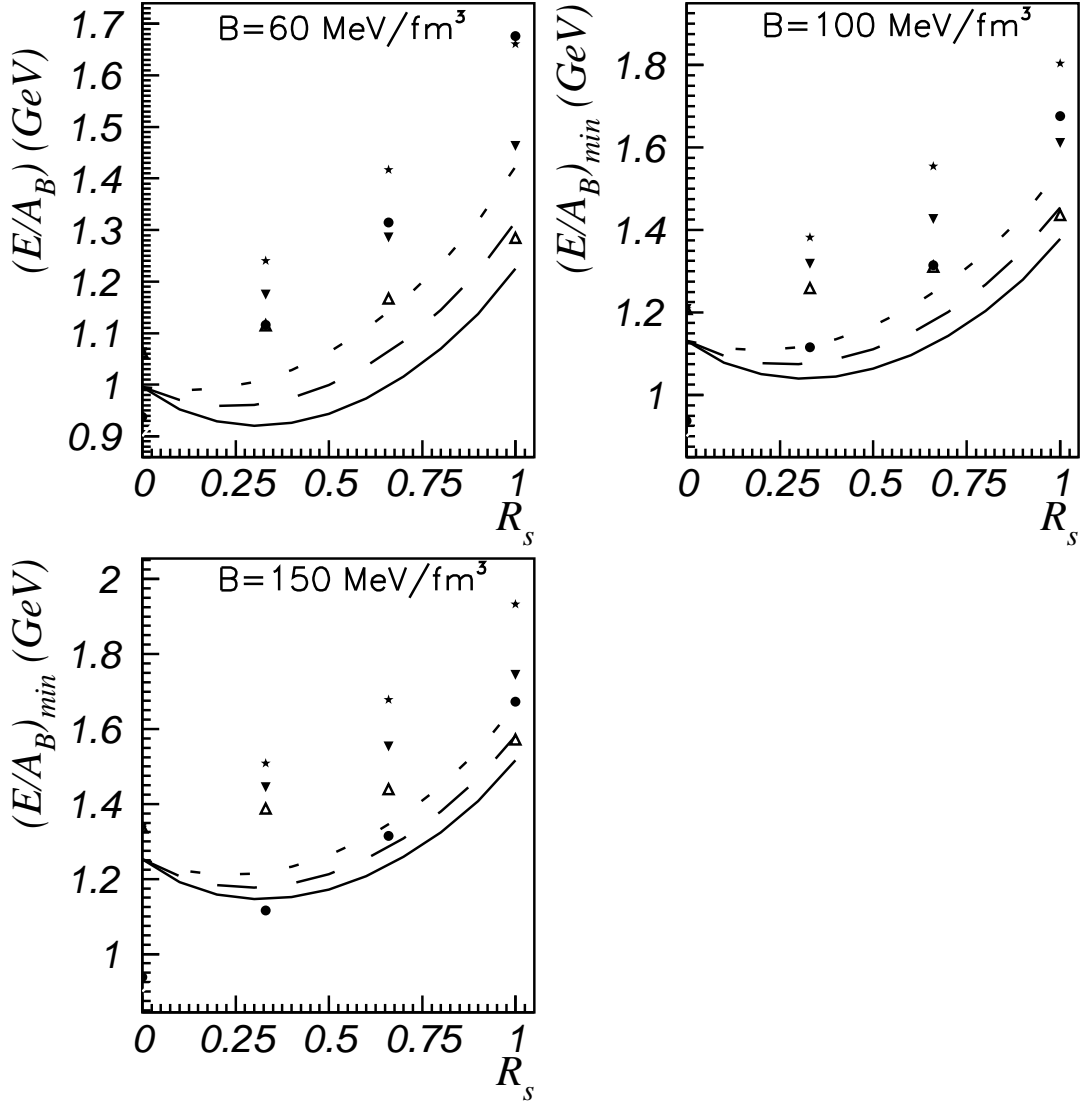,width=1\textwidth}}
\end{center}
\vskip -0.8cm
\caption{Minimal energy per baryon number in the MIT bag model, including the 
OGE potential with $\alpha_s=0.5$, as a function of the strangeness fraction 
$R_s=\rho_s/\rho$. The continuous line corresponds to $m_s=100$~MeV, the 
dashed line to $m_s=200$~MeV and the dotted line to $m_s=300$~MeV. 
Full circles represent the experimental masses, the other points refer to the masses evaluated in the model, with $m_s=100$~MeV (open triangles), $m_s=200$~MeV (full triangles), $m_s=300$~MeV (stars), respectively.
}
\label{fig4}
\end{figure}
%

%
\begin{figure}[htb]
\begin{center}
\mbox{\epsfig{file=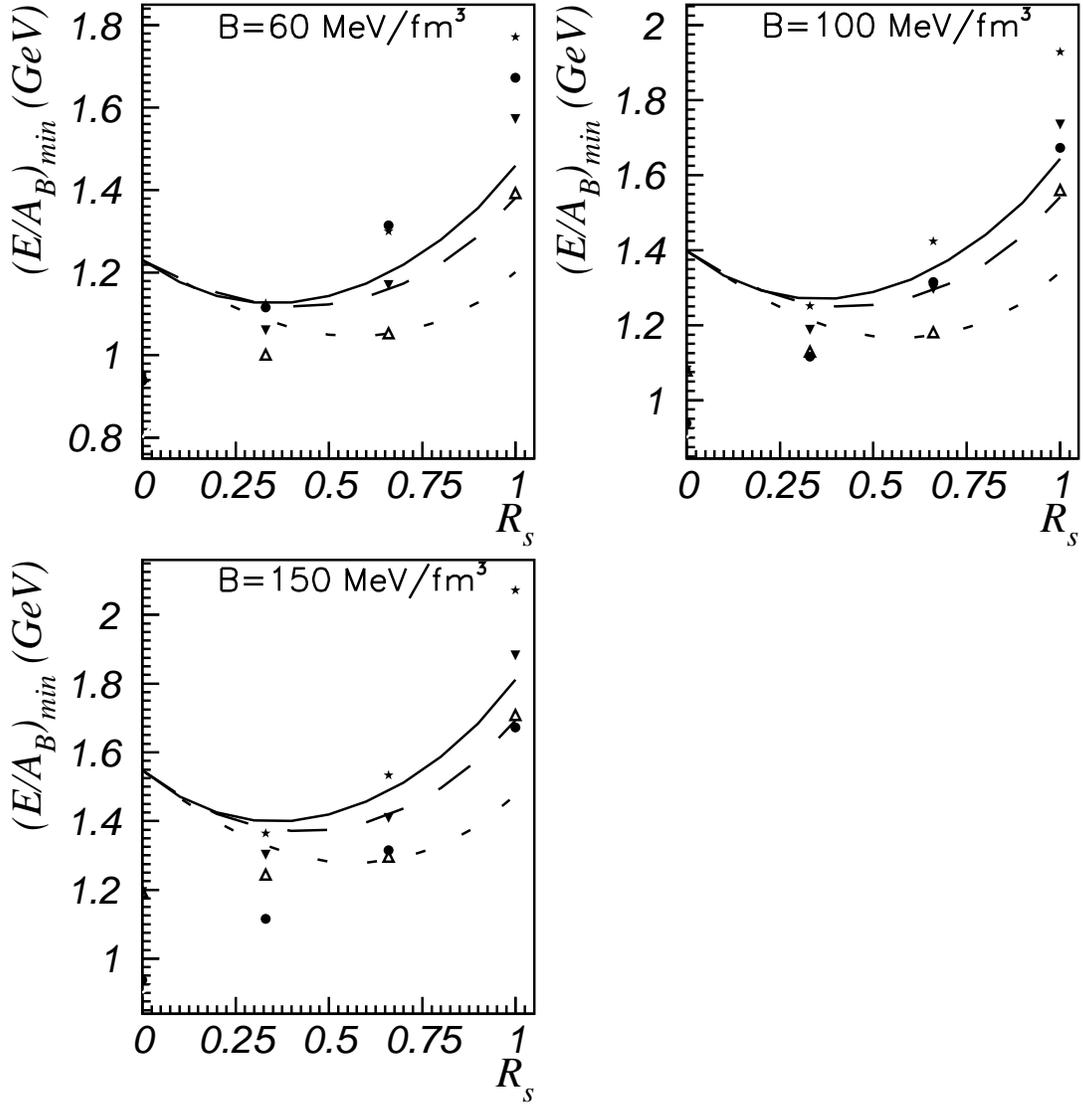,width=1\textwidth}}
\end{center}
\vskip -0.8cm
\caption{The same as in Fig.~\ref{fig4}, but for $\alpha_s=2.2$.
}
\label{fig5}
\end{figure}

\begin{figure}
\begin{center}
\mbox{\epsfig{file=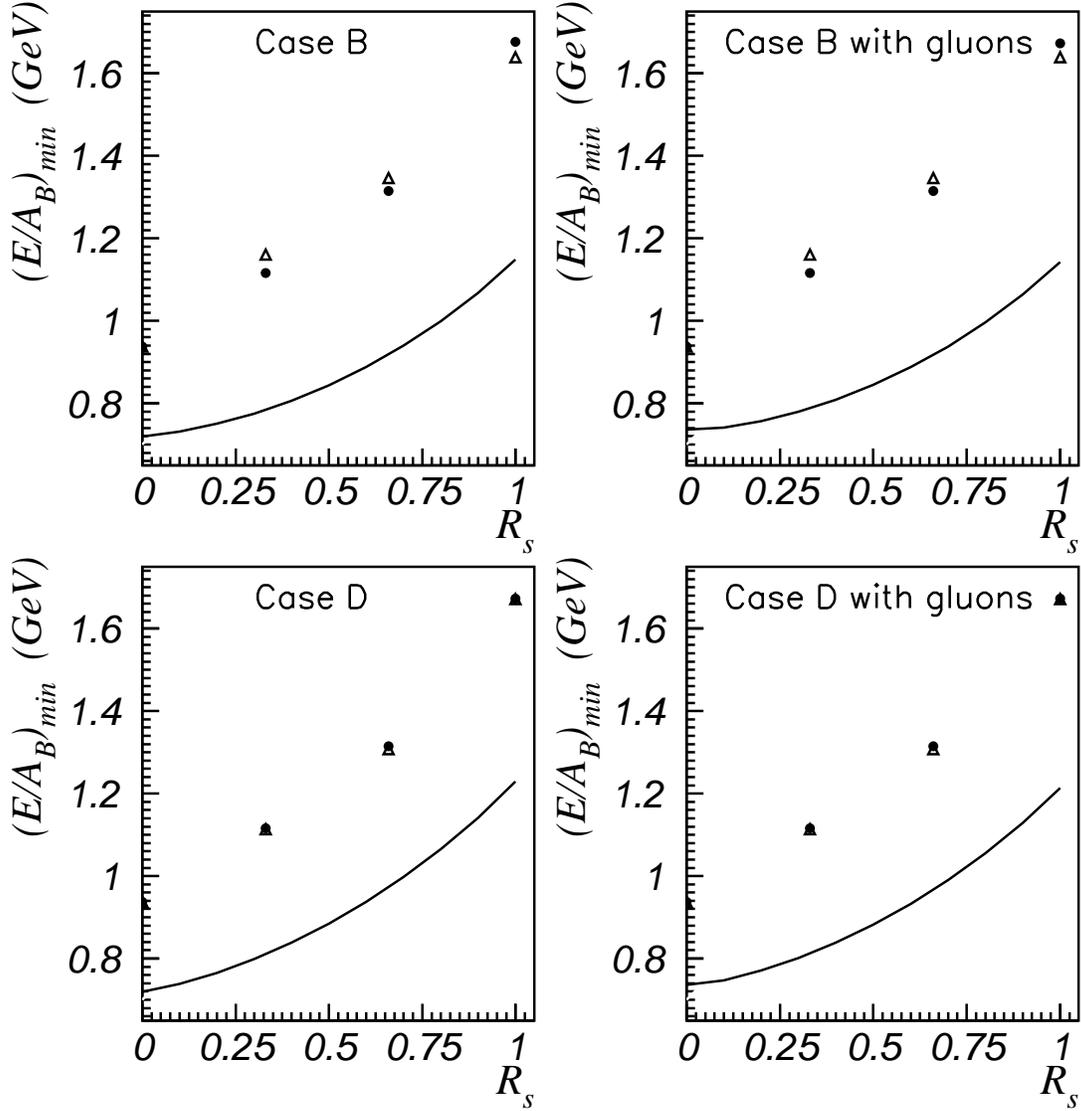,width=\textwidth}}
\end{center}
\vskip -0.8cm
\caption{Minimal energy per baryon number in the CDM, as a function of 
$R_s=\rho_s/\rho$, 
for the cases B and D with and without gluons. Full circles are the 
experimental hyperon masses, while triangular dots are the masses calculated 
in Ref.~\cite{Aoki91}.
}
\label{fig6}
\end{figure}
\begin{figure}
\begin{center}
\mbox{\epsfig{file=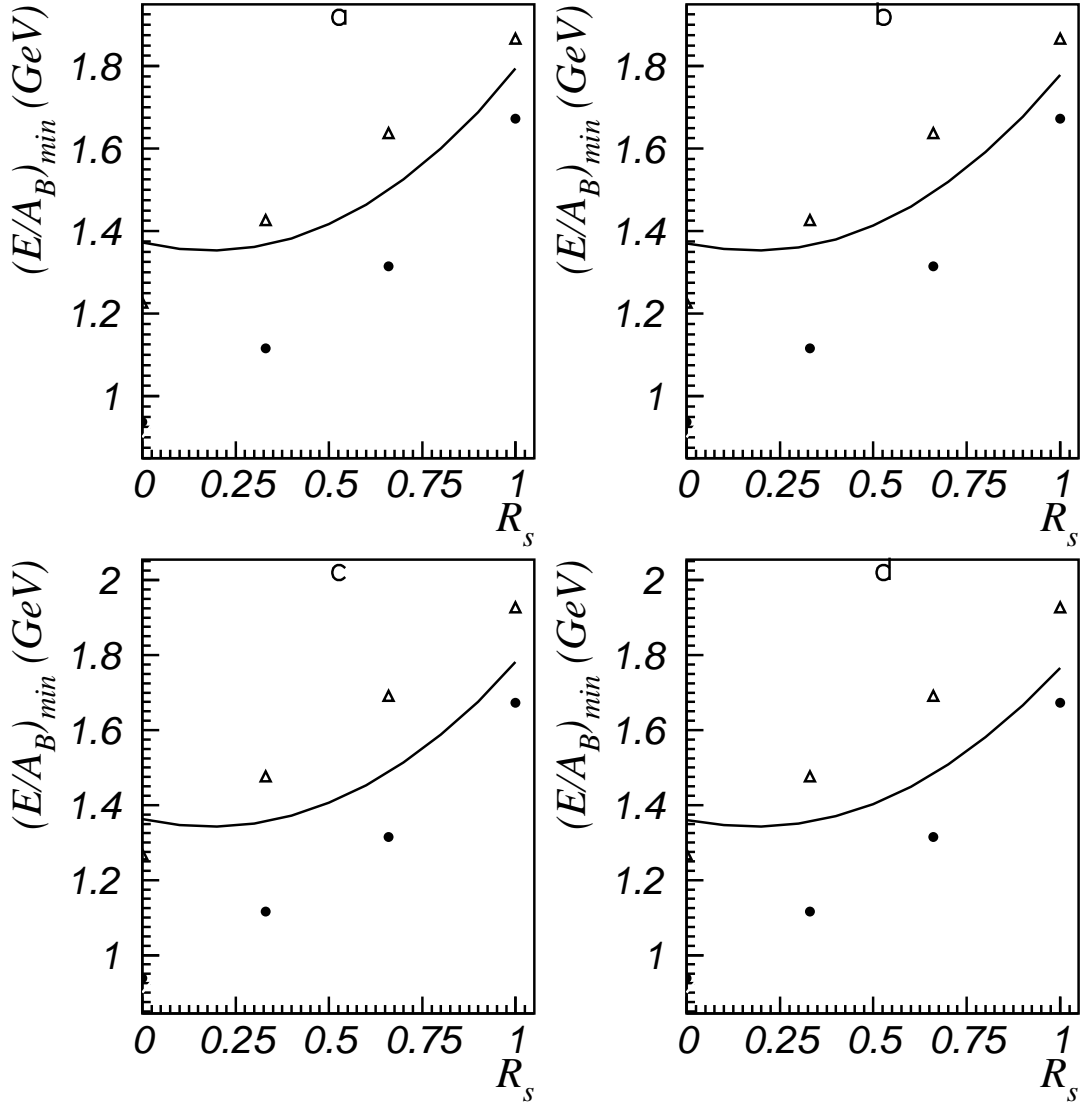,width=\textwidth}} 
\end{center}
\vskip -0.8cm
\caption{Minimal energy per baryon number as a function of $R_s=\rho_s/\rho$ for the Single Minimum version of the CDM.
The various panels correspond to: (a) parameter set I without gluons, 
(b) parameter set I with gluons, (c) parameter set II 
without gluons and (d) parameter set II with gluons. Full circles are the experimental baryon masses, while triangular dots are the masses calculated in Ref.
\cite{McGovern}}
\label{fig9}
\end{figure} 

\begin{figure}
\begin{center}
\mbox{\epsfig{file=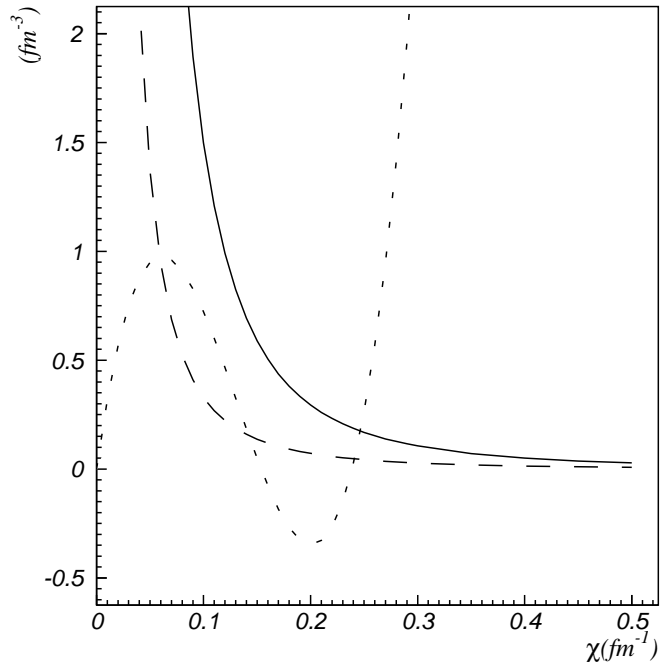,width=.6\textwidth}} 
\end{center}
\vskip -0.8cm
\caption{Solutions of the field equation for the scalar field, for two different values of $\rho$: the dotted line corresponds to d$U_{DM}/$d$\chi$, the dashed and continuous lines correspond to the r.h.s.of eq.~(\ref{scalmfdm}), assuming $\rho=0.01$ fm$^{-3}$ (dashed line) and $\rho=0.05$ fm$^{-3}$ (continuous line), with fixed $R_s=0.05$ in both cases.}
\label{fig10}
\end{figure} 
\begin{figure}
\begin{center}
\mbox{\epsfig{file=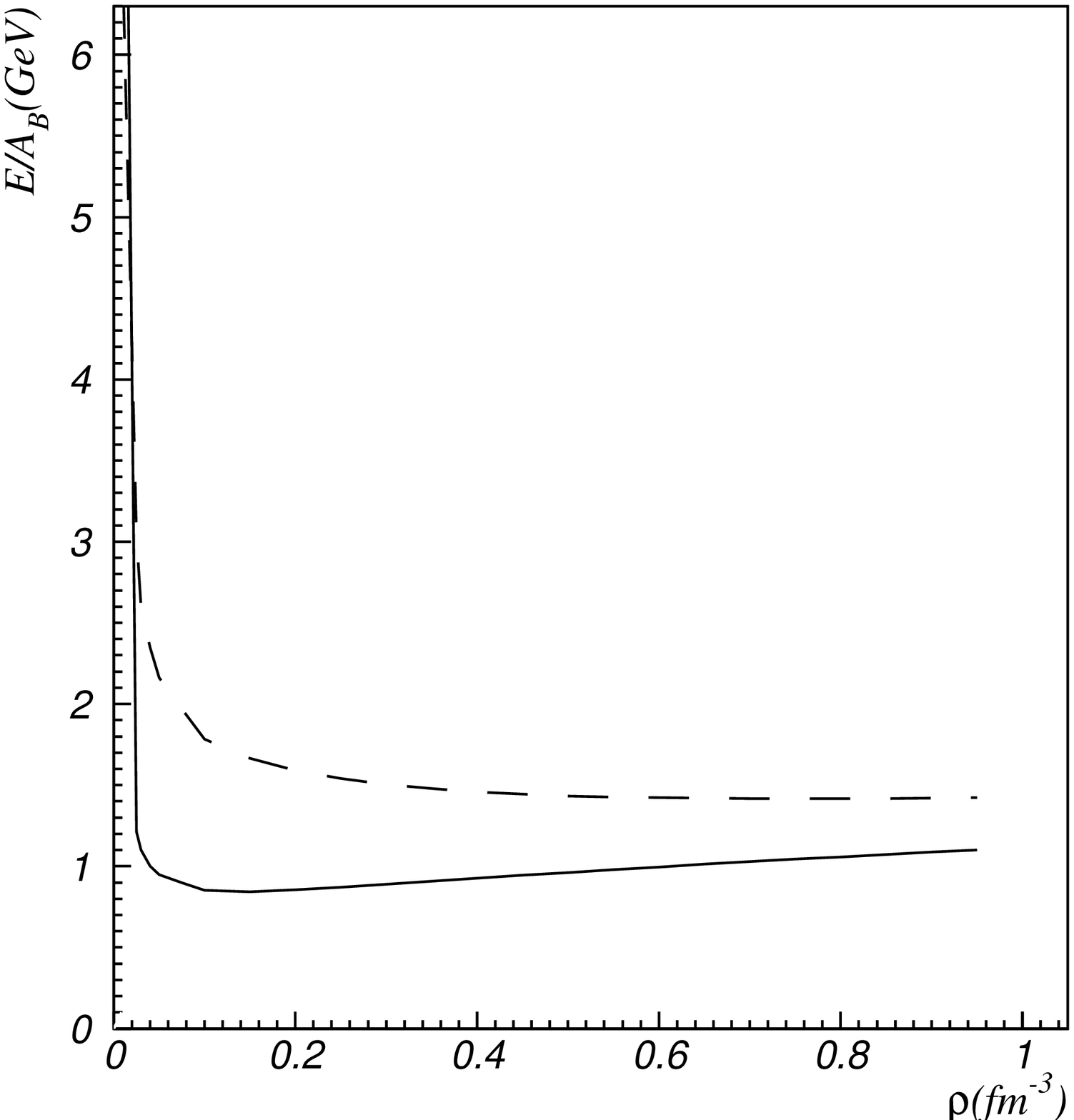,width=.6\textwidth}} 
\end{center}
\vskip -0.8cm
\caption{Total energy per baryon number as a function of $\rho$ for $R_s=0.5$: the solid line corresponds to the curve calculated with $\bar\chi$ near to the value of the relative minimum of the potential, the dashed line corresponds to $\bar\chi$ near to the absolute minimum of the potential.}
\label{fig11}
\end{figure}


\begin{thebibliography}{99}
\bibitem{Greiner88}
C. Greiner, D.-H. Rischke, H. St\"ocker and P. Koch, Z. Physik {\bf C 38},
283 (1988).
\bibitem{Greiner88b}
C. Greiner, D.-H. Rischke, H. St\"ocker and P. Koch,
Phys. Rev. {\bf D 38}, 2797 (1988).

\bibitem{expvecchi}
T.~Abbott {\it et al.}  [E-802 Collaboration],
Phys.\ Rev.\ Lett.\  {\bf 64} (1990) 847;Phys.\ Rev. Lett. {\bf 66} (1991)
 1567; O.~Hansen, Comm. Nucl. Part. Phys. {\bf 20} (1992) 1; S.~Abatzis 
{\it et al.}  [WA85 Collaboration],
Phys.\ Lett.{\bf B 316} (1993) 615; Nucl.\ Phys. {\bf A 566} (1994) 225C; 
J.B.~Kinson {\it et al.} [WA85 Collaboration], Nucl. Phys.{\bf A 544}, (1992)
 321C;
E.~Andersen {\it et al.}  [NA36 Collaboration],
Phys.\ Lett.{\bf B 294} (1992) 127; Phys.\ Rev.{\bf C 46}, (1992) 727; 
Phys.\ Lett.{\bf B 327} (1994) 433;
M.~Gazdzicki  [NA35 Collaboration.], Nucl.\ Phys.{\bf A 566} (1994) 503C;
T.~Alber {\it et al.}  [NA35 Collaboration.], Z.\ Phys.{\bf C 64} (1994) 195.

\bibitem{expnuovi}
S.~V.~Afanasev {\it et al.}  [NA49 Collaboration], J.\ Phys. {\bf G27} 
(2001) 367; D.~Varga  [the NA49 Collaboration], hep-ex/0105035;
M.~Abreu {\it et al.}  [NA50 Collaboration], Nucl.\ Phys.{\bf A 663} (2000) 
721; D.~Rohrich  [NA49 Collaboration], Nucl.\ Phys.{\bf A 663} (2000) 713.

\bibitem{Mattie89}
R. Mattiello, H. Sorge, H. St\"ocker and W. Greiner, Phys. Rev. Lett. {\bf 63},
1459 (1989).
\bibitem{Bodmer71}
A.R. Bodmer, Phys. Rev. {\bf D 4}, 1601 (1971).
\bibitem{Chin79}
S.A. Chin and A.K. Kerman, Phys. Rev. Lett. {\bf 43}. 1292 (1979).
\bibitem{Witten84}
E. Witten, Phys. Rev. {\bf D 30}, 272 (1984).
\bibitem{Dover93}
C.B. Dover, Production of strange clusters in relativistic heavy ion 
collisions, preprint BNL--48594, 1993, presented at HIPAGS 1993.
\bibitem{Greiner96}
C.~Greiner and J.~Schaffner, Int. J. Mod. Phys. {\bf E 5}, 239 (1996).
\bibitem{Greiner:1998jy}
C.~Greiner and J.~Schaffner-Bielich,
To be published in 'Heavy Elements and Related New Phenomena', 
ed. by R.K. Gupta and W. Greiner, World Scientific Publications, 
nucl-th/9801062.
\bibitem{Fahri84}
E. Fahri and R.L. Jaffe, Phys. Rev. {\bf D30}, 2379 (1984).
\bibitem{Schaffner97}
J.~Schaffner-Bielich, C.~Greiner, A.~Diener and H.~Stocker,
Phys.\ Rev.{\bf C 55} 3038 (1997).
\bibitem{Madsen00}
J. Madsen, Phys. Rev. Lett. {\bf 85}, 4687 (2000).
\bibitem{Greiner:1999by}
C.~Greiner,
J.\ Phys. {\bf G 25}, 389 (1999)
[hep-ph/9809268].

\bibitem{DeGrand75}
T. DeGrand, R.L. Jaffe, K. Johnson and J. Kiskis, Phys. Rev. {\bf D 12}, 
2060 (1975).
\bibitem{Thomas95}
F.M. Steffens, H.Holtmann, A.W. Thomas, Phys.Lett. {\bf B 358}, 139 (1995)

\bibitem{Satz82}
H.Satz Phys. Lett. {\bf B 113}, 245 (1982)
\bibitem{Andrea}
N. K. Glendenning, {\it Compact Stars}, 1997 Springer-Verlag, New York. 
\bibitem{Wilets}
L. Wilets, {\it Chiral Solitons}, ed. K.-F.Liu (World Scientific, 
Singapore, 1987) 362.
\bibitem{Birse90}
M.C. Birse, Prog. Part. Nucl. Phys. {\bf 25}, 1 (1990).
\bibitem{Pirner92}
H.~Pirner, Prog.\ Part.\ Nucl.\ Phys.\  {\bf 29}, 33 (1992).
\bibitem{Chanfray84}
H.~J.~Pirner, G.~Chanfray and O.~Nachtmann,
Phys.\ Lett. {\bf B 147}, 249 (1984).
\bibitem{Dodd87}
L.R. Dodd, A.G. Williams and A.W. Thomas, 
Phys. Rev. {\bf D 35}, 1040 (1987).
\bibitem{Aoki90}
N. Aoki and H. Hyuga, Nucl. Phys. {\bf A 505}, 525 (1989).
\bibitem{Aoki91}
K. Nishikawa, N. Aoki and H. Hyuga, 
Nucl. Phys. {\bf A 534}, 573 (1991).
\bibitem{McGovern}
J.A. McGovern, Nucl. Phys. {\bf A 533}, 553 (1991).
\bibitem{Barone93}
V.~Barone and A.~Drago,
Nucl.\ Phys. {\bf A 552}, 479 (1993).
\bibitem{Barone94}
V.~Barone, A.~Drago and M.~Fiolhais,
Phys.\ Lett. {\bf B 338}, 433 (1994).
\bibitem{Drago95}
A.~Drago, M.~Fiolhais and U.~Tambini,
Nucl.\ Phys. {\bf A 588}, 801 (1995).
\bibitem{Barone95}
V.~Barone and A.~Drago,
J.\ Phys. {\bf G21}, 1317 (1995).
\bibitem{Drago01}
A.~Drago and A.~Lavagno,
Phys.\ Lett. {\bf B 511}, 229 (2001).
\bibitem{Neuber93}      
T. Neuber, M. Fiolhais, K. Goeke and J.N. Urbano,
Nucl. Phys. {\bf A 560}, 909 (1993).
\bibitem{FetWal}
A.L. Fetter and J.D. Walecka, {\it Quantum Theory of Many-particle systems}, 
McGraw--Hill (1971).
\bibitem{Dodd87}
L.R. Dodd, A.G. Williams and A.W. Thomas, Phys.Rev. {\bf D 35}, 1040(1987).
\bibitem{Bro}
W. Broniowski, M.K. Banerjee and T.D. Cohen, MdDP-PP-87-035, ORO-5126-298,
unpublished
\end{thebibliography}
\end{document}